\documentclass[pra,aps,twocolumn,floatfix,bibnotes,showpacs]{revtex4}
\usepackage{epsfig}

\newcommand{\mywidehat}{}
\newcommand{\one}{1}
\newcommand{\two}{2}

\begin{document}

\title{Discrimination between non perfectly known states}

\author{Miroslav \surname{Je{\v{z}}ek}}
\email[]{mailto:jezek@optics.upol.cz}
\affiliation{Department of Optics, Palack\'{y} University,
             17. listopadu 50, 772 00 Olomouc, Czech Republic}

\date{\today}

\begin{abstract}
  Discrimination task is treated in the case of only partial prior
  information from measurements of unknown states.
  The construction of the optimal discrimination device and
  estimation of unknown states is performed simultaneously.
  A communication through a noisy quantum channel is formulated
  in terms of the proposed discrimination protocol.
\end{abstract}


\maketitle

\section{Introduction}
\label{intro}

Non-orthogonality of generic quantum states as a pure
quantum mechanical phenomenon poses a fundamental obstacle
to discrimination between them.
% to our ability to discriminate between them.
Impossibility of perfect distinguishing between non-orthogonal
quantum states limits the efficiency of many quantum information
protocols, for example communication through a noisy quantum channel.
Historically, the first investigation of quantum discrimination
was based on von Neumann measurements \cite{vonNeumann_QM}.
% It determined the error rate bound of two pure states discrimination.
It determined the error rate bound of discrimination between
two pure states.
The general Bayesian approach to distinguishing between mixed states with
non-equal prior probabilities based on minimization of cost function
and positive operator valued measurements (POVM) was developed
by Holevo and Helstrom \cite{Holevo_1973,Helstrom_QDET}.
In 1987 Ivanovic published a different model of two linear independent
pure states discrimination without errors but with
% a possible inconclusive result \cite{Ivanovic_1987}.
the occurrence of an inconclusive result \cite{Ivanovic_1987}.
Subsequently, the minimal probability of the inconclusive result was
shown by Dieks and Peres \cite{Dieks_1988,Peres_1988} to be the overlap
between the states being distinguished.
Recently, discrimination techniques have been thoroughly studied,
and analytical expressions for optimal POVMs for many quantum
discrimination problems have been derived
\cite{Jaeger_Shimony_1995,Chefles_Barnett_1997_JMO,%
Sasaki_Kato_Izutsu_Hirota_1998,%
Chefles_1998,Chefles_Barnett_1998,%
Phillips_Barnett_Pegg_1998,% Zhang_et_al_2001,%
Barnett_2001,% Chefles_2001,%
Walgate_et_al_2000,Virmani_et_al_2001}.
However analytical results cannot be easily obtained for discrimination
problems which incorporate many quantum states without a particular
symmetry and numerical methods appear efficient
\cite{Helstrom_1982,%
Jezek_Rehacek_Fiurasek_2002}.
Some discrimination devices have been realized experimentally
\cite{Huttner_et_al_1996,Barnett_Riis_1997,%
Clarke_Chefles_Barnett_Riis_2001,Clarke_et_al_2001}.
% Discrimination techniques can be important for quantum states
% comparison, classification and retrodicting too
% \cite{ Barnet, Chefles, Sasaki, ... }

So far distinguishability has been studied in the case of maximal
information on the states to be distinguished.
In this classical approach it is assumed that the sources
$H_i$ prepare physical systems, for example
electron spins or modes of light, in explicitly known quantum states
$\mywidehat\rho_i$. Discrimination consists in distinguishing
between these known states $\mywidehat\rho_i$ or equivalently sources
$H_i$ (they are called hypotheses) having a physical system from
one randomly chosen source.
The formal solution of this problem is based on the construction
of the optimal POVM that minimizes a convenient cost
function, for example the error rate or negative information
\cite{Holevo_1973,Helstrom_QDET}.
The goal of this paper is to develop a discrimination technique
applicable to the case when only partial information on
hypotheses $H_i$ is available.
Let us imagine that the quantum states $\mywidehat\rho_i$ prepared
by the sources $H_i$ are not explicitly known but we
are allowed to probe
each source before discrimination. We can ask what quantum
states are produced by these sources and what is the optimal
POVM to distinguish between them. This scenario realizes 
for example in communication through an unknown distorting quantum
channel. The hypotheses are explicitly known by transmitter party (Alice)
but not by receiver party (Bob). In the calibration stage Alice
sends physical systems in the quantum states $\mywidehat\rho'_i$
from the sources $H_i$ in given order down the channel while Bob
receives sequence of distorted states $\mywidehat\rho_i$.
He acquires some information (prior data) on hypothesis $H_i$
through a generic measurement of states $\mywidehat\rho_i$.
In communication stage, which follows, Alice sends a message
with the help of her alphabet $H_i$. Transmission is successful
if Bob is able to discriminate between hypothesis $H_i$ or
equivalently between quantum states $\mywidehat\rho_i$
with the help of prior data.

The article is organized as follows.
The standard discrimination is summarized in section \ref{std_app}.
Novel method of prior data manipulation for simultaneous estimation
of the unknown states and design of the optimal discrimination POVM
is proposed in section \ref{novel_app}. Two examples are shown
in section \ref{examples}.

\section{Standard approach}
\label{std_app}

Let us consider the following discrimination protocol.
Alice chooses in secret a quantum state from the set of two known
quantum states $|\one\rangle,\, |\two\rangle$ and sends it to Bob
who must decide which state was chosen using a POVM measurement.
For simplicity, but without the loss of generality, we will consider
the same probability for choosing $|\one\rangle$ and $|\two\rangle$.
A natural measure of Bob's failure is the error rate
${\rm ER} = (N_{\one\two} + N_{\two\one})/2N$,
where $N$ is the number of trials,
$N_{\one\two}$ is the number of Bob's wrong decisions
$|\one\rangle$ when $|\two\rangle$ was true and
$N_{\two\one}$ is the number of his wrong decisions
$|\two\rangle$ when $|\one\rangle$ was true.
Nonorthogonality of states implies the ultimate lower limit
of the error rate---the Helstrom bound \cite{Helstrom_QDET},
\begin{equation}  \label{Helstrom_bound}
  {\rm ER} \geq \frac12 \left( 1 -
  \sqrt{1 - |\langle\one|\two\rangle|^2}\right).
\end{equation}

The optimal POVM measurement
$\mywidehat\Pi_{\one},\, \mywidehat\Pi_{\two}$
for the discrimination between two mixed states
$\mywidehat\rho_{\one},\, \mywidehat\rho_{\two}$
is the one that minimizes the error rate
\begin{equation}  \label{error_rate}
  {\rm ER} = \frac12 \biggl(
  {\rm Tr}[\mywidehat\Pi_{\one} \mywidehat\rho_{\two}] +
  {\rm Tr}[\mywidehat\Pi_{\two} \mywidehat\rho_{\one}]
  \biggr)
\end{equation}
subject to the constraint
$\mywidehat\Pi_{\one} + \mywidehat\Pi_{\two} = \mywidehat\openone$,
or equivalently maximizes the functional
\begin{equation}  \label{min_error_rate_functional}
  {\rm I}\!{\rm E} =
  {\rm Tr}\left[\mywidehat\Pi_{\one} \mywidehat\rho_{\one}\right] +
  {\rm Tr}\left[\mywidehat\Pi_{\two} \mywidehat\rho_{\two}\right] -
  {\rm Tr}\left[\mywidehat\lambda
                (\mywidehat\Pi_{\one} + \mywidehat\Pi_{\two})\right],
\end{equation}
where $\mywidehat\Pi_{\one,\two}$ are now treated
as independent variables.
Variation of (\ref{min_error_rate_functional}) subject to the positivity
constraint on the operators $\mywidehat\Pi_{\one,\two}$
gives the extremal equations for the optimal POVM elements and for the
Lagrange operator multiplier $\mywidehat\lambda$,
\begin{equation}  \label{min_error_rate_eqs_I}
\begin{array}{c}
  {\displaystyle
  \mywidehat\rho_{\one} \mywidehat\Pi_{\one} =
    \mywidehat\lambda\mywidehat\Pi_{\one}, \qquad
  \mywidehat\rho_{\two} \mywidehat\Pi_{\two} =
    \mywidehat\lambda\mywidehat\Pi_{\two}, } \\
  {\displaystyle
  \mywidehat\rho_{\one} \mywidehat\Pi_{\one} +
    \mywidehat\rho_{\two} \mywidehat\Pi_{\two} = \mywidehat\lambda. }
\end{array}
\end{equation}
% or the equivalent set of conditions \cite{Helstrom_QDET},
% \begin{equation}  \label{min_error_rate_eqs_II}
% \begin{array}{c}
%   \mywidehat\lambda - \mywidehat\rho_{\rm A} \geq \mywidehat0, \qquad
%   \mywidehat\lambda - \mywidehat\rho_{\rm B} \geq \mywidehat0, \\
%   \mywidehat\rho_{\rm A}\mywidehat\Pi_{\rm A} +
%   \mywidehat\rho_{\rm B}\mywidehat\Pi_{\rm B} = \mywidehat\lambda.
% \end{array}
% \end{equation}

For example, solving extremal equations (\ref{min_error_rate_eqs_I})
% or (\ref{min_error_rate_eqs_II}) respectively 
for two known pure states of qubit
$
 |\psi_{\one,\two}\rangle =
 \cos\alpha \, |+z\rangle \pm \sin\alpha \, |-z\rangle
$
or corresponding mixed states
\begin{equation}  \label{states_parametrization}
  \mywidehat\rho_{\one,\two} = \left( \matrix{
    \cos^2\alpha              & \pm d\cos\alpha\sin\alpha \cr
    \pm d\cos\alpha\sin\alpha & \sin^2\alpha
  } \right),
\end{equation}
parametrized by $\alpha \in [0,\pi/4]$ and $d \in [0,1]$,
yields the optimal POVM and ultimate error rate,
\begin{equation}  \label{Helstrom_bound_for_mixed_states}
\begin{array}{c}
  {\displaystyle
  \mywidehat\Pi_{\one,\two} =
    \frac12 \left( \matrix{1 & \pm 1 \cr \pm 1 & 1} \right) =
    |\pm x\rangle\langle\pm x|, } \\
  {\displaystyle
  {\rm ER} = \frac12 (1 - d \sin2\alpha). }
\end{array}
\end{equation}

\section{Discrimination from partial information}
\label{novel_app}

In the previous section we assumed that Bob knows the states
$\mywidehat\rho_{\one,\two}$ explicitly and thus
he can use this maximal prior information for designing the
optimal discrimination POVM.
However in real communication through an unknown noisy channel
Bob does not know these states perfectly.
Therefore a calibration stage needs to be performed first.
In this stage Alice transmits a given number of physical
systems in state $\mywidehat\rho'_{\one}$ from the
known source $H_{\one}$ through communication channel and
Bob applies a prior POVM measurement $\mywidehat\pi_{k}$ to
the unknown received state $\mywidehat\rho_{\one}$.
He obtains real noisy data $f_{\one k}$ that approximate
true probabilities $p_{\one k} =
{\rm Tr}\left[ \mywidehat\rho_{\one} \mywidehat\pi_{k} \right]$.
The same calibration is done for source $H_{\two}$.
The prior information obtained by Bob during the calibration stage
is equivalent to the explicit knowledge of the states
$\mywidehat\rho_{\one,\two}$ only in the non-physical case
of tomographically complete POVM $\mywidehat\pi_{k}$
and infinitely large number of transmitted states.
The most simple and straightforward solution of this modified
discrimination problem is obvious. Bob can estimate the states
$\mywidehat\rho_{\one}$, $\mywidehat\rho_{\two}$
from the prior data $f_{\one k}$, $f_{\two k}$ and then
solve the equations
% (\ref{min_error_rate_eqs_I},\ref{min_error_rate_eqs_II})
(\ref{min_error_rate_eqs_I})
in order to obtain the optimal POVM $\mywidehat\Pi_{\one,\two}$.
Unfortunately, succession of these two steps does not
represent the optimal strategy.
If the data are not sufficient for accurate reconstruction 
of the states the deviations can be propagated to discrimination step.
Another important problem arises from solution of the extremal equations
derived from non-symmetrical mixed states obtained as a result
of some numerical state reconstruction technique
\cite{Vogel_Risken_1989,%
%% Smithey_Beck_Raymer_Faridani_1993,%
%% Kurtsiefer_Pfau_Mlynek_1997,%
Leonhardt_MQSL,%
Hradil_1997,%
%% Hradil_Summhammer_Rauch_1999,%
Banaszek_D'Ariano_Paris_Sacchi_2000,%
%% Hradil_Summhammer_Badurek_Rauch_2000,%
Rehacek_Hradil_Jezek_2001}.

\begin{figure}[t]
  \epsfxsize=6.0cm
  \centerline{\epsfbox{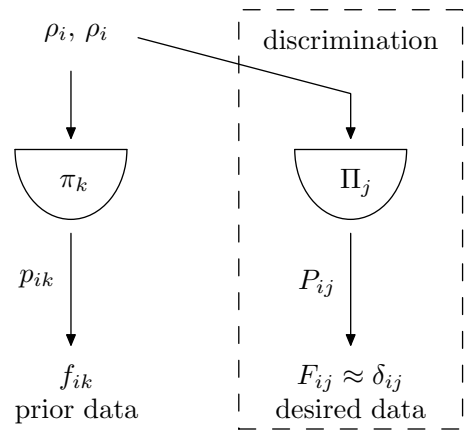}}
  \caption{Scheme of the experimental setup used for prior
           measurement of quantum states in calibration stage
           and design of the optimal discrimination POVM.}
  \label{setup_figure}
\end{figure}

Now we will formulate the modified discrimination protocol
in the general case of $I$ sources $H_i$, $i = 1,\ldots,\,I$,
and will shown information-based solution of this problem.
Let us start with $I$ unknown mixed states $\mywidehat\rho_i$
produced by the sources $H_i$ and propagated through a noisy channel.
We observe these states in the calibration stage by prior POVM
$\mywidehat\pi_k$, $k = 1,\ldots,\,K$, and obtain prior data
$f_{ik}$, $\sum_k f_{ik} = 1$, that approximate the true probabilities
\begin{equation}  \label{prior_probabilities}
  p_{ik} = {\rm Tr}\left[ \mywidehat\rho_i \mywidehat\pi_k \right].
\end{equation}
In the discrimination stage the states $\mywidehat\rho_i$ are
examined in virtual sense by designed discrimination POVM
$\mywidehat\Pi_j$, $j = 1,\ldots,\,J$. If we actually
performed this measurement we would obtain frequencies
$F_{ij}$, $\sum_j F_{ij} = 1$, sampling probabilities
\begin{equation}  \label{virtual_probabilities}
  P_{ij} = {\rm Tr}\left[ \mywidehat\rho_i \mywidehat\Pi_j \right].
\end{equation}
For simplicity, let us consider the same number of states
$\mywidehat\rho_i$ and designed POVM elements $\mywidehat\Pi_j$,
$I=J$. In the case of more POVM elements than discriminated states
the proposed method would generalize the Ivanovic--Dieks--Peres
\cite{Ivanovic_1987,Dieks_1988,Peres_1988} unambiguous
discrimination or combination of both methods,
not the pure Helstrom scheme.
% \cite{Helstrom_QDET}.
For optimal POVM the sum of diagonal frequencies
$\sum_i F_{ii}$ should be maximal and the error rate
$\sum_{ij} F_{ij} - \sum_i F_{ii}$
should be minimal. Thus the desired discrimination POVM
should yield data $F_{ij}$ as close to Kronecker's
delta $\delta_{ij}$ as possible.
% Measurement scheme is plotted in Fig. \ref{setup_figure}.
This is schematically shown in Fig. \ref{setup_figure}.

In the case of the finite number of states used for calibration
the equalities $f_{ik} = p_{ik}$ do not hold exactly.
Similarly the equalities $F_{ij} = P_{ij} = \delta_{ij}$
are not satisfied because the perfect discrimination between
non-orthogonal states is not possible.
These equalities are satisfied only approximately and thus
cannot be directly used for the estimation of the states
$\mywidehat\rho_i$ and design of the optimal POVM $\mywidehat\Pi_j$.
% In the limit of the infinite number of states used in calibration stage
% and essential discrimination POVM the equalities $f_{ik} = p_{ik}$
% and $F_{ij} = P_{ij} = \delta_{ij}$ are valid.
% Unfortunately both limiting cases are non-physical, equalities
% are fulfilled only approximately and cannot be used for
% estimation  of states $\mywidehat\rho_i$ and design of optimal POVM
% $\mywidehat\Pi_j$.
Therefore equality between data $f_{ik}$, $F_{ij}$ and probabilities
$p_{ik}$, $P_{ij}$ must be handled in weaker -- statistical -- manner.
We will construct a functional
${\rm I}\!{\rm F}[f_{ik},$ $F_{ij},$
$p_{ik}(\mywidehat\rho_i),$ $P_{ij}(\mywidehat\rho_i,\mywidehat\Pi_j)]$
that will serve as a measure of the distance between the data and the
probabilities. It should be maximal for states
$\mywidehat\rho_i$ which produce prior probabilities $p_{ik}$ close to
prior data $f_{ik}$ as possible and simultaneously
it should be maximal for and POVM $\mywidehat\Pi_j$ which produces
probabilities $P_{ij}$ as close to the desired data $F_{ij} = \delta_{ij}$
as possible.
There are many convex and additive measures of
distance between probability distributions.
We will choose the one that provides
the most simple extremal equations---the log likelihood functional,
\begin{equation}  \label{likelihood}
  \log {\rm I}\!{\rm L} [\mywidehat\rho_i,\mywidehat\Pi_j] =
  \sum_{ij} \delta_{ij} \ln P_{ij} +
  \sum_{ik} f_{ik} \ln p_{ik}.
\end{equation}
Log likelihood as a measure of distance between quantum mechanical
objects and real data
arises from non-normalized multinomial distribution
\begin{equation}  \label{multinomial}
  {\rm I}\!{\rm L} [\mywidehat\rho_i,\mywidehat\Pi_j] \approx
  \prod_{ijk} P_{ij}^{\delta_{ij}} p_{ik}^{f_{ik}} =
  \prod_{ik} P_{ii} \, p_{ik}^{f_{ik}}.
\end{equation}
and has unique properties from the physical \cite{Hradil_Summhammer_2000}
as well as mathematical \cite{Rao_SI} point of view.
% There is also tight connection between maximum likelihood inference
% and Bayesian approach with uniform priors \cite{Jones_1994}.

Variation of the log likelihood functional (\ref{likelihood}),
\begin{equation}  \label{likelihood_ext}
  \log {\rm I}\!{\rm L}
  [\mywidehat\rho_i + \mywidehat{X}_i, \mywidehat\Pi_j + \mywidehat{Y}_j] -
  \log {\rm I}\!{\rm L} [\mywidehat\rho_i, \mywidehat\Pi_j] = 0,
\end{equation}
for all $\mywidehat{X}_i$, $\mywidehat{Y}_j$, that yields extremal equations
for the states $\mywidehat\rho_i$
and the optimal POVM $\mywidehat\Pi_j$ must preserve
trace normalization of the states,
${\rm Tr}[\mywidehat\rho_i] = 1$,
and completeness of the POVM, $\sum_j \mywidehat\Pi_j = \mywidehat\openone$.
This necessary constraints can be incorporated with
the help of undetermined
Lagrange multipliers. The functional to be maximized becomes
\begin{eqnarray}
  {\rm I}\!{\rm F}[\mywidehat\rho_i,\mywidehat\Pi_j] & = &
  \sum_{ij} \delta_{ij} \ln P_{ij} +
  \sum_{ik} f_{ik} \ln p_{ik} - \nonumber \\
  & &
  \sum_i \mu_i {\rm Tr}[\mywidehat\rho_i] -
  {\rm Tr}[\mywidehat\lambda \sum_j \mywidehat\Pi_j],
  \label{likelihood_c}
\end{eqnarray}
where $\mywidehat\rho_i$ and $\mywidehat\Pi_j$ are the independent
variables.
The estimated unknown states $\mywidehat\rho_i^{\rm est}$ and the designed
optimal discrimination POVM $\mywidehat\Pi_j^{\rm opt}$ which maximize
the functional (\ref{likelihood_c}) are solutions of the
extremal equations
\begin{equation}  \label{Max_likelihood_eqs}
  \mu_i^{-2}
  \mywidehat{R}_i \mywidehat\rho_i \mywidehat{R}_i = \mywidehat\rho_i,
  \qquad
  \mywidehat\lambda^{-1}
  \mywidehat{S}_j \mywidehat\Pi_j \mywidehat{S}_j
  \mywidehat\lambda^{-1} = \mywidehat\Pi_j,
\end{equation}
where
\begin{eqnarray}
  &&
  \mywidehat{R}_i = \frac{1}{P_{ii}} \mywidehat\Pi_i +
    \sum_k \frac{f_{ik}}{p_{ik}} \mywidehat\pi_k,
  \qquad
  \mywidehat{S}_j = \frac{1}{P_{jj}} \mywidehat\rho_j,
  \label{kernels} \\
  &&
  \mu_i =
  \left( {\rm Tr}\left[
    \mywidehat{R}_i \mywidehat\rho_i \mywidehat{R}_i
  \right] \right)^{\frac12},
  \quad
  \mywidehat\lambda = \biggl(
    \sum_j \mywidehat{S}_j \mywidehat\Pi_j \mywidehat{S}_j
  \biggr)^{\frac12}.
  \label{lagrange_multipliers}
\end{eqnarray}
Notice that $\mywidehat\lambda$ is positive definite operator
and $\mu_i$ are positive numbers.
The set of nonlinear operator equations
(\ref{Max_likelihood_eqs}--\ref{lagrange_multipliers})
can be solved by means of repeated iterations,
\begin{equation}  \label{ext_iteration}
\begin{array}{l}
  (\mu_i^{(n)})^{-2}
  \mywidehat{R}_i^{(n)} \mywidehat\rho_i^{(n)} \mywidehat{R}_i^{(n)}
  = \mywidehat\rho_i^{(n+1)}, \\
  (\mywidehat\lambda^{(n)})^{-1}
  \mywidehat{S}_j^{(n)} \mywidehat\Pi_j^{(n)} \mywidehat{S}_j^{(n)}
  (\mywidehat\lambda^{(n)})^{-1} = \mywidehat\Pi_j^{(n+1)}.
\end{array}
\end{equation}
The convexity of the log likelihood functional guarantees that
these equations have only one global maximum
or a plateau of global maxima.
As initial iterations we can choose
maximally mixed states and POVM elements
$\mywidehat\Pi_j^{(0)} = \frac1J \mywidehat\openone$.
Let us note that when the true states $\mywidehat\rho_i$ are mixed
the estimated states $\mywidehat\rho_i^{\rm est}$
% which fulfill
% extremal equations (\ref{Max_likelihood_eqs}--\ref{lagrange_multipliers})
are biased towards pure states due to the simultaneous construction
of the discrimination POVM $\mywidehat\Pi_j$.
The reason is that, roughly speaking,
the rays are what is important for discrimination,
and pure states can be discriminated easier than mixed ones.
In another words, the proposed method finds the
discrimination POVM $\mywidehat\Pi_j^{\rm opt}$
{\em optimal in the sense of maximum likelihood\/},
and simultaneously reconstructs rays corresponding
to the unknown states.

\section{Examples}
\label{examples}

Here we illustrate our procedure on discrimination 
between two qubit states $\mywidehat\rho_{\one,\two}$, such as
two spin states of an electron or two polarization states
of a photon. We perform  % carry out
numerical simulations of the prior measurements and we find
subsequently the states $\mywidehat\rho_{\one,\two}^{\rm est}$
and determined the optimal POVM $\mywidehat\Pi_{\one,\two}^{\rm opt}$
via iterative solution of the extremal equations
(\ref{Max_likelihood_eqs}--\ref{lagrange_multipliers}).
In our simulations we assume the class of states
(\ref{states_parametrization}) and study the dependence
of the error rate (\ref{error_rate}) on angle $\alpha$
that controls the overlap between the two discriminated
states $\mywidehat\rho_{\one,\two}$.
The error rate (\ref{error_rate}) computed from the true
states $\mywidehat\rho_{\one,\two}$ and the optimal POVM
$\mywidehat\Pi_{\one,\two}^{\rm opt}$ is used for
evaluating the quality of the designed POVM
$\mywidehat\Pi_{\one,\two}^{\rm opt}$.
Thus the optimality of the discrimination POVM
$\mywidehat\Pi_{\one,\two}^{\rm opt}$ obtained with the help
of the proposed method is quantified by the {\em standard\/}
widely used measure of discrimination success.

\begin{figure}[ht]
  \epsfxsize=8.5cm 
  \centerline{\hspace*{-0.5cm}\epsfbox{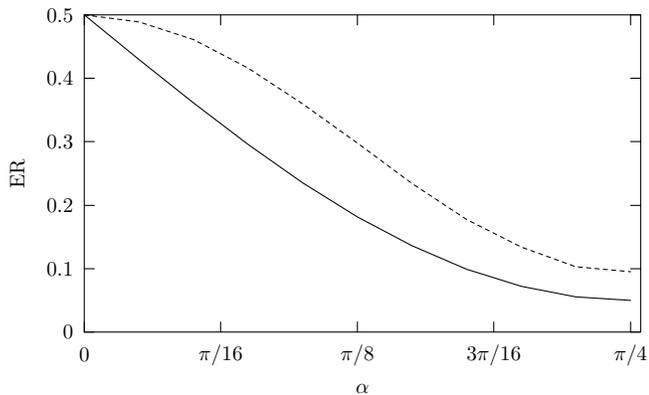}}
  \caption{Error rate of the optimal discrimination between
  two symmetrical mixed states using partial prior information
  from the tomographically incomplete measurement (dashed), see text.
  Prior data $f$ fluctuate around their theoretical values $p$
  with standard deviation $\sqrt{p(1-p)}/\sqrt{N},$ $N=1000$.
  Error rate is limited by Helstrom bound (solid).}
  \label{error_rate_1_figure}
\end{figure}

In the first example we consider two symmetrical mixed
states (\ref{states_parametrization}) with $d=0.9$
on the Bob's side of the communication channel.
The prior POVM $\mywidehat\pi_k$ consists of
projective measurements in $x$ and $y$ directions,
each made on $N=1000$ physical systems in states
$\mywidehat\rho_{\one,\two}$.
This represents the calibration stage.
Observation
$\{\mywidehat\pi_{x},\, \mywidehat\pi_{-x},\,
   \mywidehat\pi_{y},\, \mywidehat\pi_{-y}\}$
is tomographically incomplete and thus not sufficient
for estimation of mixed states. Nevertheless, this measurement
yields some information on the states $\mywidehat\rho_{\one,\two}$
and, discrimination with the error rate less than $50\%$ is possible.
The error rate (\ref{error_rate}),
$\frac12 (
{\rm Tr}[\mywidehat\Pi_{\one}^{\rm opt} \mywidehat\rho_{\two}] +
{\rm Tr}[\mywidehat\Pi_{\two}^{\rm opt} \mywidehat\rho_{\one}])$,
of the optimal discrimination POVM
$\mywidehat\Pi_{\one,\two}^{\rm opt}$
is shown in Fig.~\ref{error_rate_1_figure}.

\begin{figure}[bh]
  \epsfxsize=8.5cm 
  \centerline{\hspace*{-0.5cm}\epsfbox{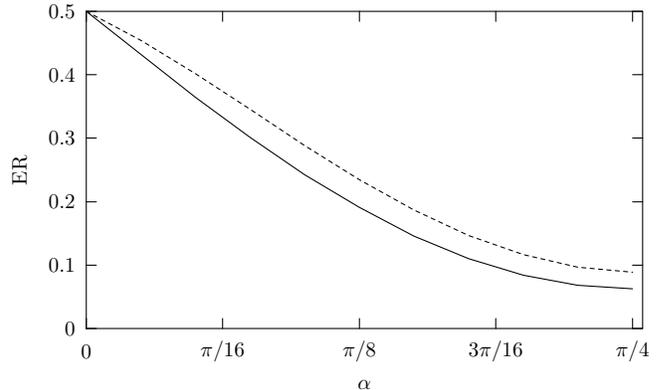}}
  \caption{The error rate of the optimal discrimination between
  the pure and the mixed state using the partial prior information
  from the tomographically complete measurement (dashed), see text.
  Prior data $f$ fluctuate around their theoretical values $p$
  with standard deviation $\sqrt{p(1-p)}/\sqrt{N},$ $N=10$.
  Error rate is limited by Helstrom bound (solid).}
  \label{error_rate_2_figure}
\end{figure}

In the second example we discriminate between the pure
state $\mywidehat\rho_{\one} =
|\psi_{\one}\rangle \langle\psi_{\one}|$,
$d=1$, and the mixed state $\mywidehat\rho_{\two}$
(\ref{states_parametrization}) with $d=3/4$,
having prior data from the tomographically complete observation
$\{\mywidehat\pi_{x},\, \mywidehat\pi_{-x},\,
   \mywidehat\pi_{y},\, \mywidehat\pi_{-y},
   \mywidehat\pi_{z},\, \mywidehat\pi_{-z}\}$.
Each of these projective measurements in $x$, $y$ and $z$
directions are done with only $N=10$ physical systems
in states $\mywidehat\rho_{\one}$ and $\mywidehat\rho_{\two}$.
The optimal POVM $\mywidehat\Pi_{\one,\two}^{\rm opt}$
can be found with the help of (\ref{ext_iteration}).
The corresponding error rate is shown
in Fig.~\ref{error_rate_2_figure}.
In the case of an infinitely large number of physical systems
used for prior measurement in the calibration stage the prior
data $f_{ik}$ attain the theoretical probabilities $p_{ik}$
and the error rate approaches the bound
(\ref{Helstrom_bound_for_mixed_states}),
${\rm ER} = \frac{1}{2} (1 - \frac{1+d}{2}\sin{2\alpha}) =
\frac{1}{2} (1 - \frac{7}{8}\sin{2\alpha})$, as is shown
in Fig.~\ref{error_rate_2_figure} by solid line.

\section{Conclusion}
\label{summary}

Simultaneous reconstruction of quantum states and design
of optimal quantum discrimination POVM have been presented.
% Simultaneous quantum states reconstruction and optimal
% quantum discrimination POVM design have been presented.
Observation on finite samples of physical systems in unknown
quantum states in the calibration stage provides the only
prior information used for discrimination between these
quantum states.
The system of the extremal equations
(\ref{Max_likelihood_eqs}--\ref{lagrange_multipliers})
fully determines the unknown states and the optimal POVM
for their discrimination.
These nonlinear operator equations can be solved by means
of repeated iterations.
This has been demonstrated on particular examples and
verified in many other cases.
Typically several hundreds or thousands of iterations are
needed for $10^{-16}$ accuracy.
In the case of finite samples of physical systems used for
the calibration the Helstrom bound represents a lower limit
of the achievable error rate. For tomographically complete prior
observation the error rate can come very close to this bound
even if the prior measurements are carried out on a small
number of states.
This means that in practice Alice and Bob need only a few
transmitted quantum states for establishing nearly optimal
discrimination device. The calibration stage thus can be very
short and almost all the states reserved for communication can be
used up for transmission of useful data from Alice to Bob.
Of course the ratio of the transmitted calibration states to the
states used for communication can be adjusted in order to achieve
either transmission with maximal speed for the given error rate
or transmission with minimal error rate for the given speed.
These and others optimization problems can be formulated
as a modifications of the presented quantum discrimination protocol.

%% \section*{Acknowledgements}

This work was supported by Grant LN00A015 and
by Research Project CEZ: J14/98: 153100009 ``Wave and particle optics''
of the Czech Ministry of Education.
The author would like to thank Radim Filip,
Jarom{\'{\i}}r Fiur{\'{a}}{\v{s}}ek, Zden{\v{e}}k Hradil and
Jaroslav {\v{R}}eh{\'{a}}{\v{c}}ek for very helpful discussions
and Computer Science Department of Palack{\'{y}} University
for computer time.

%% BIBLIOGRAPHY from discrimination.bib
%% \bibliography{discrimination}

\end{document}